\documentclass[onecolumn,showpacs,amsmath,amssymb,showkeys]{revtex4}

\usepackage{graphicx}
\usepackage{epsfig}
\usepackage{bm}

\begin{document}

\title{On Schr\"{o}dinger equation  with potential
$U= - \alpha r^{-1} + \beta r + kr^{2} $ \\
and the bi-confluent Heun functions theory }

\author{E. Ovsiyuk}

\affiliation{Mozyr State Pedagogical University\\
28 Studencheskaya Str., Mozyr, 247760, Gomel region, Belarus}

\author{M. Amirfachrian}

\affiliation{Institute of Physics, National Academy of Sciences of
Belarus\\  68 Nezalezhnasci av., Minsk, 220072, Belarus }

\author{O. Veko}

\affiliation{Mozyr State Pedagogical University\\
28 Studencheskaya Str., Mozyr, 247760, Gomel region, Belarus}


\begin{abstract}

It is shown that Schr\"{o}dinger equation with  combination of
three potentials $U= - \alpha r^{-1} + \beta r + kr^{2} $,
Coulomb, linear and harmonic, the potential often used to describe
quarkonium,
 is  reduced to a bi-confluent Heun differential equation.
 The method to construct  its solutions in the form of polynomials is developed,
 however with additional constraints
  in four parameters of the model, $ \alpha, \beta ,  k, l $.
  The energy spectrum looks as a modified combination
  of oscillator and Coulomb parts.

\end{abstract}



\maketitle

\section{Introduction}

Gauge theories of the strong interactions  suggest that the
coupling between quarks is weak at short distances but becomes
very strong at large distances. This 'explains' the paradox where
quarks appear to behave as quasi-free particles within hadrons but
cannot be liberated from the hadrons. On the basis of these
arguments it was conjectured that heavy quarks would move
nonrelativistically within hadrons. Thus bound states of a heavy
quark and antiquark should be a hadronic analogue of the
positronium system of a bound electron and positron.

This so-called quarkonium system might then be interpreted
according to the familiar rules of nonrelativistic quantum
mechanics using a potential to describe the interquark force (see
\cite{1}, \cite{2} and references therein).

 At very small
distances, the potential is expected to take a form like the
Coulomb force, corresponding to the exchange of a single massless
gluon. At very large distances, a linear confining potential or
something more complex  seem to be appropriate. We will use a
combination of three well-known potentials
\begin{eqnarray}
U= - {\alpha \over r } + \beta r + kr^{2}
\nonumber
\end{eqnarray}

\noindent that has  some advantages for analytical treatment of
the quarkonium  problem in terms of Heun functions (this class of special functions being
next extension to hypergeometric functions has become of primary importance in many
physical problems   \cite{3}--\cite{32}).

\section{Schr\"{o}dinger equation}

In Minkowski space, parameterized  by spherical coordinate
\begin{eqnarray}
dS^{2} = c^{2} dt^{2} -  d  r^{2}  - r^{2} (d
\theta^2+  \sin^{2} \theta d \phi^2  ) \; ,
\nonumber
\end{eqnarray}

\noindent  the  Schr\"{o}dinger equation for an arbitrary  spherical
potential has the form
\begin{eqnarray}
i\hbar{\partial \over \partial t} \Psi=
  \left [ -{\hbar^{2}\over 2 M }  \left ( {1\over r^{2}}{\partial
\over \partial r} r^{2}{\partial \over
\partial r} - {\hat{l}^{\; 2} \over r^{2} }     \right )
 +  U( r)   \; \right ]  \Psi \; .
\label{2.1}
\end{eqnarray}

\noindent After separation of the variables, $\Psi = e^{-{i E
t\over \hbar}} \; Y_{lm}(\theta, \phi) \; R(r)  \; , $ we get
\begin{eqnarray}
E\;  R =
  \left [ -{\hbar^{2}\over 2 M } \left ( {1\over r^{2}}{ d
\over d  r} r^{2}  {d  \over d r} - {l(l+1) \over r^{2}  }
\right   )
 + U (r)  \; \right ]  R \; .
 \label{2.2}
\end{eqnarray}

For a potential specified by (we assume the Coulomb attraction, so $\alpha > 0$)
\begin{eqnarray}
U=  - {\alpha \over r}  + \beta r + kr^{2}
\label{2.3}
\end{eqnarray}

\noindent  eq.    (\ref{2.2}) reads
\begin{eqnarray}
  \left [   { d^{2}
\over d  r^{2} } +{2 \over r} {d  \over d r}   - {l(l+1) \over r^{2}
}   + {2M \over \hbar^{2}} \left (  E + {\alpha \over r}  - \beta r - kr^{2} \right )
  \right ]  R  = 0 \; . \label{2.4a}
\end{eqnarray}

We may simplify notation by using (instead of (\ref{2.4a}))
more short form
\begin{eqnarray}
  \left (   { d^{2}
\over d  r^{2} } +{2 \over r} {d  \over d r}     +   2 \epsilon  + {\alpha \over r}  - {l(l+1) \over r^{2}}   - \beta r - kr^{2}
  \right )  R  = 0 \; \label{2.4b}
\end{eqnarray}

\noindent
and further with the substitution $R(r) = r^{-1} f(r)$ we get
\begin{eqnarray}
  \left (   { d^{2}
\over d  r^{2} }     +   2 \epsilon  + {\alpha \over r}  - {l(l+1) \over r^{2}}   - \beta r - kr^{2}
  \right )  f  = 0 \; . \label{2.4c}
\end{eqnarray}

Let us see behavior  of the curve
\begin{eqnarray}
P^{2}(r)= 2 \epsilon  + {\alpha \over r}  - {l(l+1) \over r^{2}}   - \beta r - kr^{2} \; ,
\nonumber\\
P^{2} (r \sim 0 ) \sim  -{l(l+1) \over r^{2}} \sim - \infty\;, \qquad
P^{2} (r \sim \infty) \sim -kr^{2}   - \infty\,.
\nonumber
\end{eqnarray}

To proceed further, let us examine the classical turning points --  the roots of the equation
\begin{eqnarray}
-kr^{4} - \beta r^{3} + 2\epsilon r^{2} + \alpha r - l(l+1) =
-k (r-r_{1}) (r-r_{2})(r-r_{3})(r-r_{4})  =0 \; .
\label{2.5a}
\end{eqnarray}

From (\ref{2.5a}) it follows
\begin{eqnarray}
-{ \beta  \over  k} = r_{1} + r_{2} + r_{3} + r_{4} \; ,
\nonumber\\
-{ 2\epsilon \over k} =  r_{1}r_{2} + r_{1}r_{3} + r_{1} r_{4} + r_{2}r_{3} + r_{2}r_{4} + r_{3}r_{4}\;,
\nonumber\\
{\alpha  \over k} =  r_{2}r_{3} r_{4} + r_{1}r_{3} r_{4} +  r_{1}r_{2} r_{4} + r_{1}r_{2} r_{3}   \; ,
\nonumber\\
{l(l+1) \over k} =  r_{1} r_{2} r_{3} r_{4} \; .
\label{2.5b}
\end{eqnarray}

There exist the possibility  when two roots are  negative  and two other are   positive
(see Fig. 1).

\vspace{+3mm}

 \unitlength=0.75 mm
\begin{picture}(160,100)(-90,-60)
\special{em:linewidth 0.4pt} \linethickness{0.4pt}

\put(-60,0){\vector(+1,0){120}}     \put(+60,-5){$r $}
\put(0,-30,0){\vector(0,+1){60}}   \put(+5,+30){$P^{2}(r) $}

\put(+40,-0.5){\circle*{2}}  \put(+39,-9){$r_{4}$}
\put(+10,-0.5){\circle*{2}}  \put(+10,-9){$r_{3}$}

\put(10,-0.5){\line(+1,0){30} }
\put(10,-0.8){\line(+1,0){30} }
\put(10,-0.3){\line(+1,0){30} }

\put(-40,-0.5){\circle*{2}}  \put(-39,+5){$r_{1}$}
\put(-10,-0.5){\circle*{2}}  \put(-10,+5){$r_{2}$}

\put(+2.5,-20){\circle*{1}}
\put(+3.5,-15){\circle*{1}}
\put(+4.6,-10){\circle*{1}}
\put(+6.5,-5){\circle*{1}}
\put(+13,+4){\circle*{1}}
\put(+15,+8){\circle*{1}}
\put(+17,+10){\circle*{1}}
\put(+21,+11.5){\circle*{1}}
\put(+23,+12){\circle*{1}}
\put(+25,+11.5){\circle*{1}}
\put(+27,+10){\circle*{1}}
\put(+30,+8){\circle*{1}}
\put(+32,+6.5){\circle*{1}}
\put(+35,+4){\circle*{1}}
\put(+38,+2){\circle*{1}}

\put(+42,-3){\circle*{1}}
\put(+44,-5){\circle*{1}}
\put(+46,-9){\circle*{1}}
\put(+48,-13){\circle*{1}}
\put(+50,-16){\circle*{1}}
\put(+52,-20){\circle*{1}}
\end{picture}

\vspace{-20mm}

\begin{center}
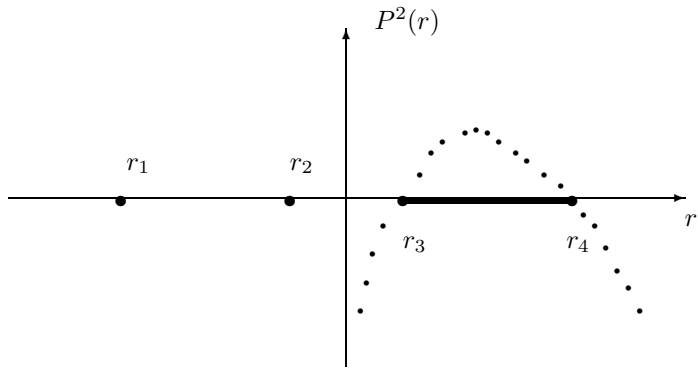
 FIG.  1:   Finite classical motion at  $r \in [ r_{3}, r_{4}]$
\end{center}

From (\ref{2.5a}) it follows two systems:
\begin{eqnarray}
-{ 2\epsilon \over k} =  r_{1}r_{2} + r_{1} (r_{3} +  r_{4} )
+ r_{2} (r_{3} + r_{4})  + r_{3}r_{4}\; ,
\nonumber\\
{\alpha  \over k} =  r_{1} \; r_{3} r_{4}  + r_{2} \; r_{3} r_{4}  +  r_{1}r_{2} (r_{4} +  r_{3} )  \; ,
\nonumber\end{eqnarray}
and
\begin{eqnarray}
 r_{1} + r_{2}  = -{ \beta  \over  k}  -  r_{3} - r_{4} \; ,
\nonumber\\
  r_{1} r_{2}  = {l(l+1) \over k \; r_{3} r_{4}  }  \; .
\nonumber
\end{eqnarray}

Making the change of variables
$y=iK r$ ($ K = k^{1/4}$):
\begin{eqnarray}
{d^{2}R\over dy^{2}}+{2\over y}\,{dR\over dy}-\left({2\epsilon\over K^{2}}+{i \alpha\over K y}+{l(l+1)\over y^{2}}+{i \beta y\over K^{3}}+y^{2}\right)R=0\,
\label{2.6}
\end{eqnarray}

\noindent and using the substitution
\begin{eqnarray}
R=y^{A}\,e^{By}\,e^{Cy^{2}}F(y) \; ,
\label{2.7}
\end{eqnarray}

 \noindent
 we get
\begin{eqnarray}
{d^{2}F\over dy^{2}}+\left[{2(A+1)\over y}+4Cy+2B\right]\,{dF\over dy}+\left[(4C^{2}-1)\,y^{2}+\left(4BC-{i\,\beta\over K^{3}}\right)\,y+\right.
\nonumber\\
\left.+4AC+B^{2}-{2 \epsilon\over K^{2}}+6C+{A^{2}+A-l(l+1)\over y^{2}}+{2ABK+2BK-i \alpha\over y\,K}\right]F=0\,.
\label{2.8}
\end{eqnarray}

With the choice of special values
\begin{eqnarray}
A  =   +l \; , \; - ( l +1) \, ,
\qquad  C= \pm {1\over 2}\, , \qquad
B= \pm {i\beta\over 2\,K^{3}}\,, \
\label{2.9a}
\end{eqnarray}

\noindent
take those  related to possible bound states
\begin{eqnarray}
A  =   +l , \qquad y^{A}  \sim  r^{l} \; ;
\nonumber\\
C=  + {1\over 2}\, , \qquad e^{Cy^{2}} = e^{-K^{2} r^{2} / 2} \; ;
\nonumber\\
B= + {i\beta\over 2\,K^{3}}\,, \qquad e^{By} =e^{-\beta r / 2K^{2}} \;
;
\label{2.9b}
\end{eqnarray}

\noindent eq. (\ref{2.8}) becomes simpler
\begin{eqnarray}
{d^{2}F\over dy^{2}}+\left[{2(A+1)\over y}+4Cy+2B\right]\,{dF\over dy}+
\nonumber\\
+\left[4AC+B^{2}-{2 \epsilon\over K^{2}}+6C+{2ABK+2BK-i \alpha\over y\,K}\right]F=0\,.
\label{2.10}
\end{eqnarray}

It is convenient to turn to the variable
\begin{eqnarray}
z = {y \over  i}  = K r \; ;\nonumber
\end{eqnarray}
then eq. (\ref{2.10}) reads (remember that  $A =+ l$ и $ C = + 1/2$)
\begin{eqnarray}
{d^{2}F\over d z^{2}}+ \left (  -2 z  +2iB   +{1 + (2l +1)\over z } \right ) {dF\over d z}+
\nonumber\\
+
\left ( - 2 - (2l  +1) + ({2 \epsilon\over K^{2}}   - B^{2} )  +  { -(-2iB)(l +1)  + \alpha / K \over   z } \right)
F=0\, ,
\label{2.11}
\end{eqnarray}

 which can be recognized as a biconfluent Heun equation for
 functions
$H(a, b, c, d, z)$
\begin{eqnarray}
{d^{2}H\over dz^{2}}+\left(-2z-b+{1+a\over z}\right)\,{dH\over dz}+\left(-2-a+c+
{-b (a +1)/2- d/2 \over z}\right)\,H=0
\label{2.12}
\end{eqnarray}

\noindent with parameters
\begin{eqnarray}
a=2l +1 \,, \qquad
b=-2iB =  {\beta\over K^{3}} \,, \qquad
c=  {2\epsilon\over K^{2}}  + {\beta^{2} \over  4 K^{6}}  \,, \qquad
d= - {2\,\alpha\over K}\,.
\label{2.13}
\end{eqnarray}

Let us present  solutions of eq.  (\ref{2.12}) as a series
\begin{eqnarray}
H (z) = 1 + c_{1} z + c_{2} z^{2} + c_{3} z^{3} + ... = \sum _{n=0}^{\infty}
c_{n}z^{n} \; ,
\nonumber\\
H' (z) =  c_{1}  + 2 c_{2} z^{1} +   3c_{3} z^{2} + ... = \sum _{n=1}^{\infty}
nc_{n}z^{n-1} \; ,
\nonumber\\
H '' (z) =   2 \times  1 \;  c_{2} z^{0} +   3 \times 2 \; c_{3} z^{1} + ... = \sum _{n=2}^{\infty}
n(n-1)c_{n}z^{n-2} \; .
\nonumber
\end{eqnarray}

\noindent Eq.  (\ref{2.12})  gives (let $D=-b (a +1)/2- d/2$)
\begin{eqnarray}
{d^{2}H\over dz^{2}}+ \left(-2z-b+{1+a\over z}\right)\,{dH\over dz}+\left(-2-a+c+
{D \over z}\right)\,H=0 \; ;
\nonumber
\end{eqnarray}

\noindent that is
\begin{eqnarray}
\sum _{n=2}^{\infty}
n(n-1)c_{n}z^{n-2} +
\nonumber\\
+ (-2z-b+{1+a\over z} ) \sum _{n=1}^{\infty}
nc_{n}z^{n-1} +
\nonumber\\
+
(-2-a+c+
{D \over z} ) \sum _{n=0}^{\infty}
c_{n}z^{n} = 0 \; .
\label{2.14}
\end{eqnarray}

From (\ref{2.14}) it follows
\begin{eqnarray}
\sum _{n=2}^{\infty} n(n-1)c_{n}z^{n-2} -
  \sum _{n=1}^{\infty} 2nc_{n}z^{n} -
  \sum _{n=1}^{\infty} b nc_{n}z^{n-1} +
\nonumber\\
+  \sum _{n=1}^{\infty} (1+a) nc_{n}z^{n-2} +
  \sum _{n=0}^{\infty} (-2-a+c) c_{n}z^{n}
+
 \sum _{n=0}^{\infty} D c_{n}z^{n-1}
\nonumber
\end{eqnarray}

\noindent
or
\begin{eqnarray}
\sum _{N=0}^{\infty} (N+2)(N+1) c_{N+2}z^{N} -
   \sum _{N=1}^{\infty} 2Nc_{N}z^{N} -
 \sum _{N=0}^{\infty} b (N+1) c_{N+1}z^{N} +
\nonumber\\
+  \sum _{N=-1}^{\infty} (1+a) (N+2)c_{N+2}z^{N} +
  \sum _{N=0}^{\infty} (-2-a+c) c_{N}z^{N}
+
 \sum _{N=-1}^{\infty} D c_{N+1}z^{N} \; .
\nonumber
\end{eqnarray}

Collecting  similar terms
\begin{eqnarray}
2  c_{2} +
\sum _{N=1}^{\infty} (N+2)(N+1) c_{N+2}z^{N} -
\nonumber\\
-   \sum _{N=1}^{\infty} 2Nc_{N}z^{N} -
   b  c_{1}  -  \sum _{N=1}^{\infty} b (N+1) c_{N+1}z^{N} +
\nonumber\\
+   (1+a) c_{1}z^{-1}
+  (1+a) 2c_{2}
+  \sum _{N=1}^{\infty} (1+a) (N+2)c_{N+2}z^{N}  +
\nonumber\\
+
  (-2-a+c) c_{0}  +  \sum _{N=1}^{\infty} (-2-a+c) c_{N}z^{N} +
\nonumber\\
+
 D c_{0}z^{-1} +   D c_{1}
 + \sum _{N=1}^{\infty} D c_{N+1}z^{N} = 0
\nonumber
\end{eqnarray}

\noindent
we get
\begin{eqnarray}
\left [   \; (1+a) c_{1} +
 D c_{0}  \right ] \; z^{-1}  +
\nonumber\\
+ \left [\; 2  c_{2} -   b  c_{1}   +  (1+a) 2c_{2} +
  (-2-a+c) c_{0} +    D c_{1}  \; \right ] +
\nonumber\\
 +
\sum _{N=1}^{\infty}  [\; (N+2)(N+1) c_{N+2} -
   2Nc_{N} -
   b (N+1) c_{N+1}+
\nonumber\\
+   (1+a) (N+2)c_{N+2}  +
  (-2-a+c) c_{N}  +
  D c_{N+1}\; ] \; z^{N} = 0\,.
\nonumber
\end{eqnarray}

Thus we arrive at the recurrent  relations for series coefficients
\begin{eqnarray}
c_{1} = -
 {D  \over  (1+a) } \; c_{0} \; , \qquad
   c_{2} =    {
  (2+a-c) c_{0} +   ( b-D)c_{1}  \over   (a+2) 2}  \; ,
  \nonumber\\
  c_{N+2}   =
  {
    (2N +2+a -c)\;  c_{N} + [ - D  +   b (N+1)]\;  c_{N+1} \over  (N+2)(a+ N+2) }
  \; , \qquad N = 1,2, ...
\label{2.15}
\end{eqnarray}

From this, after simple change in notation, we obtain
\begin{eqnarray}
c_{1} = -
 {D  \over  (1+a) } \; c_{0} \; , \qquad
   c_{2} =    {
  (2+a-c) c_{0} +   ( b-D)c_{1}  \over   (a+2) 2}  \; ,
  \nonumber\\
  c_{n+1}   =
  {
    (2n +a -c)\;  c_{n-1} + ( - D  +   b n)\;  c_{n} \over  (n+1)( a+ n +1) }
  \; , \qquad n = 2,3, 4,  ...
 \label{2.16}
\end{eqnarray}

Remember that
\begin{eqnarray}
a=2l +1 \,, \qquad
b= {\beta\over K^{3}} \,, \qquad d= - {2\,\alpha\over K}\,\qquad
c=  {2\epsilon\over K^{2}}  + {b^{2} \over  4 }  \,,
\nonumber
\end{eqnarray}

\noindent
we obtain
\begin{eqnarray}
-D=   b (l+1)  - {\alpha \over K}  \; , \qquad
b - D =
 b (l+2)   - {\alpha \over K} \;,
\nonumber\\
2b - D =   b(l+3 )   - {\alpha \over K} \;,\;\;
 . . . \;\;
bn - D =
b (l+n+1)  - {\alpha \over K}  \; ;
\label{2.17}
\end{eqnarray}

\noindent note that the energy  parameter does not enter relations in (2.17), instead it is presented
only in the parameter $c$
\begin{eqnarray}
c=  {2\epsilon\over K^{2}}  + {b^{2} \over  4 } \;.
\label{2.18}
\end{eqnarray}

Recursive relation  can be rewritten in the form
\begin{eqnarray}
c_{1} = -
 {D  \over  2 (2l+2) } \; c_{0} \; ,
\nonumber\\
   c_{2} =    {
  [ \; (2(l+1) + (1 -c ) \; ]\; c_{0} +   ( b-D)c_{1}  \over   2 \; [ (2 (l+1) + 1 ] }  \; ,
 \nonumber\\
  c_{3}   =
  {
    [\;  2 (l+2) + (1  -c )\; ] \;  c_{1} + (2 b  - D  )\;  c_{2} \over  3 \; [ 2 (l+1) +2 ]  }   \; ,
 \nonumber\\
 c_{4}   =
  {
    [ \; 2 (3 +l) + (1  -c )\; ]\;  c_{2} + (   3 b  - D    )\;  c_{3} \over  4\;
    [\; 2 (l+1) + 3 ]  }
  \; ,
\nonumber\\
  c_{5}   =
  {
    [ \; 2 (4 +l) + (1  -c )\; ]\;  c_{3} + (   4 b  - D    )\;  c_{4} \over  5\;
    [\; 2 (l+1) + 4 ]  }
  \; ,
  \nonumber
\end{eqnarray}
. . . . . . . . . . . . . . . . . . . . . . . . . . . . . . . . . . . . . . . . . .  . . . . . . . . .  . . . . . . . . .
\begin{eqnarray}
  c_{n+1}   =
  {
    [ \; 2 (n +l) + (1  -c )\; ]\;  c_{n-1} + (   n b  - D    )\;  c_{n} \over  (n+1)\;
    [\; 2 (l+1) + n ]  }
  \;  ...
\label{2.19}
\end{eqnarray}

Ont can try to obtain polynomial solutions.
The first simplest possibility  can be realized by adding two constraints:
\begin{eqnarray}
c_{1}= 0, \qquad  c_{2} = 0
\label{2.20a}
\end{eqnarray}

\noindent
which gives
\begin{eqnarray}
D= 0 \qquad \Longrightarrow \qquad  {\beta  (l+1) \over K^{2}}  = \alpha  \; ,
\nonumber\\
c = (2(l+1) + 1  \qquad \Longrightarrow \qquad  {2\epsilon \over K^{2}} + {b^{2} \over 4} = 2(l+1) +1 \; .
\label{2.20b}
\end{eqnarray}

Result (\ref{2.20b}) may be rewritten as follows

\underline{the case $n=0$}
\begin{eqnarray}
  \epsilon  = K^{2}  (l+ 1 + { 1\over 2} ) -  {1\over 8 }  ({\alpha \over l+1} )^{2} \;,
  \qquad \mbox{where} \qquad  { \alpha \over l+1} = {\beta \over K^{2}} \; .
\label{2.20c}
\end{eqnarray}

Let us consider the case $n=1$ which is realized by adding two constraints
\begin{eqnarray}
    c_{2} = 0\; , \qquad
   [\; (2(l+1) + (1 -c ) \; ]\; c_{0} +   ( b-D)c_{1}  =0 \; ,
\nonumber\\
  c_{3}   =0\;, \qquad
 2 (l+2) + (1  -c ) =0   \; ,
\label{2.21a}
\end{eqnarray}

\noindent which leads to
\begin{eqnarray}
  c =   2 (l+2) + 1\,,
\nonumber\\
 2   +   ( b-D) {D  \over  2 (2l+2) }   =0\,.
\label{2.21b}
\end{eqnarray}

In explicit form they are
\begin{eqnarray}
{2\epsilon   \over K^{2}}  +  {b^{2} \over  4 }   =  2(l+2)  + 1     \;,
\label{2.22a}
\end{eqnarray}

\noindent
and
\begin{eqnarray}
4(2l+2)  - [-{\alpha \over K} + b (l+2)] [-{\alpha \over K}  + b (l+1)]  =0\; .
\label{2.22b}
\end{eqnarray}

\noindent
Eq. (\ref{2.22b}) is  a 2-order equation with respect to $b$
\begin{eqnarray}
 b^{2} (l+1)(l+2)  - b {\alpha \over K} (2l+3) - 4(2l+2) +{\alpha^{2} \over K^{2}}    = 0
\nonumber
\end{eqnarray}

\noindent
with its solution
\begin{eqnarray}
b   = +   {\alpha \over K} {(l+3/2) \over
 (l+1)(l+2) } \pm  \left [ ({\alpha \over   K})^{2}   \left (
 {(l+3/2) \over
 (l+1) (l+2) }  \right )^{2}  +   {4(2l+2) - \alpha^{2}/ K^{2}   \over (l+1)(l+2)} \right ] ^{1/2}
\nonumber
\end{eqnarray}

\noindent or
\begin{eqnarray}
b   = +   {\alpha \over K} {(l+3/2) \over
 (l+1)(l+2) } \pm  \sqrt{  ({\alpha \over   K})^{2}
 {1 /4\over
 (l+1)^{2} (l+2)^{2}}   +   {8  \over (l+2)} } \; .
\label{2.22c}
\end{eqnarray}

 Let us consider
several first  coefficients of the bi-confluent Heun series:
\begin{eqnarray}
c_{1} = -  {D  \over  2 (2l+2) } \; c_{0} \; ;
\nonumber\\
   c_{2} =    {
   (2(l+1) + (1 -c )  \over   2 \; [ \; (2 (l+1) + 1 \; ] }
  \; c_{0} +   { ( b-D)  \over   2 \; [ \; (2 (l+1) + 1 \; ] } \; c_{1}    =
 \nonumber\\
= {
   (2(l+1) + (1 -c )  \over   2 \; [ \; (2 (l+1) + 1 \; ] }
  \; c_{0} +   { ( b-D)  \over   2 \; [ \; (2 (l+1) + 1 \; ] } \; \left (
  -  {D  \over  2 (2l+2) } \; c_{0}  \right )\; ;
  \nonumber\\
  c_{3}   =
  {
      2 (l+2) + (1  -c )   \over  3 \; [ \; 2 (l+1) +2 \; ]  }  \;  c_{1} +
   { (2 b  - D  )\;   \over  3 \; [ \; 2 (l+1) +2 \; ]  }\; c_{2}   =
  \nonumber\\
=
{
      2 (l+2) + (1  -c )   \over  3 \; [ \; 2 (l+1) +2 \; ]  }  \; \left ( -  {D  \over  2 (2l+2) } \; c_{0} \right ) +
      \nonumber\\
+   { (2 b  - D  )\;   \over  3 \; [ \; 2 (l+1) +2 \; ]  }\;
   \left [
   {
   (2(l+1) + (1 -c )  \over   2 \; [ \; (2 (l+1) + 1 \; ] }
  \; c_{0} +   { ( b-D)  \over   2 \; [ \; (2 (l+1) + 1 \; ] } \; \left (
  -  {D  \over  2 (2l+2) } \; c_{0}  \right )  \right ]   \; ;
  \nonumber\\
 c_{4}   =
  {
    2 (3 +l) + (1  -c ) \over  4\;     [\; 2 (l+1) + 3 ]  }
    \;  c_{2} + { (   3 b  - D    )    \over  4\;     [\; 2 (l+1) + 3 ]  } \; \;  c_{3}
  =
\nonumber\\
=
 {
    2 (3 +l) + (1  -c ) \over  4\;     [\; 2 (l+1) + 3 ]  }
    \;  \left [
    {
   (2(l+1) + (1 -c )  \over   2 \; [ \; (2 (l+1) + 1 \; ] }
  \; c_{0} +   { ( b-D)  \over   2 \; [ \; (2 (l+1) + 1 \; ] } \; \left (
  -  {D  \over  2 (2l+2) } \; c_{0}  \right ) \right ] +
   \nonumber\\
    +
     { (   3 b  - D    )    \over  4\;     [\; 2 (l+1) + 3 ]  } \; \;
     \left \{
{
      2 (l+2) + (1  -c )   \over  3 \; [ \; 2 (l+1) +2 \; ]  }  \; \left ( -  {D  \over  2 (2l+2) } \; c_{0} \right ) +
      \right.
     \nonumber\\
\left. +   { (2 b  - D  )\;   \over  3 \; [ \; 2 (l+1) +2 \; ]  }\;
   \left [
   {
   (2(l+1) + (1 -c )  \over   2 \; [ \; (2 (l+1) + 1 \; ] }
  \; c_{0} +   { ( b-D)  \over   2 \; [ \; (2 (l+1) + 1 \; ] } \; \left (
  -  {D  \over  2 (2l+2) } \; c_{0}  \right )  \right ]    \right \} \; ;
  \nonumber
\end{eqnarray}
. . . . . . . . . . . . . . . . . . . . . . . . . . . . . . . . . . . . . . . . . .  . . . . . . . . .  . . . . . . . . .

The coefficient $c_{n}$ represents a $n$-polynomial with respect to parameter $b$.

In principle, it is easily to extend  the above polynomial-based  approach to general case.
Indeed, let it be
\begin{eqnarray}
c_{n+1}=0\; , \qquad c_{n+2}= 0 \; ,
\label{2.23a}
\end{eqnarray}

\noindent which gives
\begin{eqnarray}
 c_{n+1}=0, \qquad
    [ \; 2 (n +l) + (1  -c )\; ]\;  c_{n-1} + (   n b  - D    )\;  c_{n}=0
  \; ,
  \nonumber\\
c_{n+2}=0, \qquad
    [ \; 2 (n+1 +l) + (1  -c )\; ]\;  c_{n}  =0
  \; .
 \label{2.23b}
\end{eqnarray}

\noindent From whence it follows
\begin{eqnarray}
 (1  -c ) = - 2 (n+1 +l)\,,
\nonumber\\
    -2   c_{n-1} + (   n b  - D    )\;  c_{n}=0
  \; .
\label{2.23c}
\end{eqnarray}

The problem is  reduced  to rather complicated polynomials.
This
method  provides  us with the formula for energy levels;
however we obtain some  additional constraints
for four parameters,
$\alpha, \beta, k, l$ (in the form of $n$-polynomial). This means that we are able to construct solutions
 in the polynomial form,
 but only at some special values of  $\alpha, \beta, k, l$.

General structure of the 3-term recurrent relations can  be presented in a more short notation
\begin{eqnarray}
c_{1} = A_{0} c_{0} \; ,
\nonumber\\
c_{2} = E_{0} c_{0} + A_{1} c_{1}\; ,
\nonumber\\
c_{3} = E_{1} c_{1} + A_{2} c_{2}
\;,
\nonumber\\
c_{4} = E_{2} c_{2} + A_{3} c_{3} \; ,
\nonumber\\
c_{5} = E_{3} c_{3} + A_{4} c_{4}
\; ,
\nonumber\\
................................
\nonumber\\
c_{n+1} = E_{n-1} c_{n-1} + A_{n} c_{n} \; ,
\nonumber\\
................................
\label{2.24}
\end{eqnarray}

\noindent
where
\begin{eqnarray}
E_{n-1} = {
     2 (n +l) + (1  -c ) \over  (n+1)\;     [\; 2 (l+1) + n ]  }, \qquad
     A_{n} = { (   n b  - D    )     \over  (n+1)\;     [\; 2 (l+1) + n ]  }  \; .
 \label{2.25}
\end{eqnarray}

\section{Acknowledgements}

This  work was   supported   by the Fund for Basic Researches of Belarus
 F11M-152.

\end{document}